 \documentclass[doublespacing]{elsart}

 \usepackage{graphics}
 \usepackage{graphicx}
 \usepackage{epsfig}

\usepackage{amssymb}

\begin{document}

\begin{frontmatter}



\title{Precise Measurements of Atmospheric Muon Fluxes with the BESS Spectrometer}


\author[label1,label7]{M.~Motoki\corauthref{label11}},
\corauth[label11]{
Corresponding author.
}
\ead{motoki@awa.tohoku.ac.jp}
\author[label1]{T.~Sanuki},
\author[label1,label8]{S.~Orito},
\author[label1]{K.~Abe},
\author[label1]{K.~Anraku},
\author[label1,label9]{Y.~Asaoka},
\author[label1]{M.~Fujikawa},
\author[label1]{H.~Fuke},
\author[label1]{S.~Haino},
\author[label1]{M.~Imori},
\author[label1]{K.~Izumi},
\author[label2]{T.~Maeno},
\author[label3]{Y.~Makida},
\author[label1]{N.~Matsui},
\author[label1]{H.~Matsumoto},
\author[label1,label10]{H.~Matsunaga},
\author[label4]{J.~Mitchell},
\author[label2,label7]{T.~Mitsui},
\author[label4]{A.~Moiseev},
\author[label1]{J.~Nishimura},
\author[label2]{M.~Nozaki},
\author[label4]{J.~Ormes},
\author[label1]{T.~Saeki},
\author[label4]{M.~Sasaki},
\author[label5]{E.~S.~Seo},
\author[label1]{Y.~Shikaze},
\author[label1]{T.~Sonoda},
\author[label4]{R.~Streitmatter},
\author[label3]{J.~Suzuki},
\author[label3]{K.~Tanaka},
\author[label1]{I.~Ueda},
\author[label5]{J.~Z.~Wang},
\author[label6]{N.~Yajima},
\author[label6]{T.~Yamagami},
\author[label3]{A.~Yamamoto},
\author[label1]{Y.~Yamamoto},
\author[label2]{K.~Yamato},
\author[label3]{T.~Yoshida},
\author[label3]{K.~Yoshimura}

\address[label1]{University of Tokyo, Tokyo, 113-0033, Japan}
\address[label2]{Kobe University, Kobe, Hyogo, 657-8501, Japan}
\address[label3]{High Energy Accelerator Research Organization (KEK),
 Tsukuba, Ibaraki, 305-0801, Japan}
\address[label4]{National Aeronautics and Space Administration,
 Goddard Space Flight Center (NASA/GSFC), Greenbelt, MD, 20771, USA}
\address[label5]{University of Maryland, College Park, MD 20742, USA}
\address[label6]{The Institute of Space and Astronautical Science (ISAS), Sagamihara,
 Kanagawa, 229-8510, Japan}

\thanks[label7]{
 Present address: Research Center for Neutrino
		Science, Tohoku 
		University, Sendai, 980-8578, Japan.
}
\thanks[label8]{deceased.}
\thanks[label9]{
 Present address: ICRR, University of Tokyo, Kashiwa, Chiba,
		277-8582, Japan.
}
\thanks[label10]{
 Present address: University of Tsukuba, Ibaraki,
		305-8571, Japan.
}

\begin{abstract}

The vertical absolute fluxes of atmospheric muons and muon charge
ratio 
have been measured precisely at different geomagnetic locations
by using the BESS spectrometer.
The observations had been performed at sea level (30~m above sea level)
in Tsukuba, Japan, and at 
360~m above sea level in Lynn Lake, Canada.
The vertical cutoff rigidities in
Tsukuba~($36.2^{\circ}N$,$140.1^{\circ}E$) and in Lynn
Lake~($56.5^{\circ}N$,$101.0^{\circ}W$) are 11.4~GV and 0.4~GV,
respectively. 
 We have obtained 
vertical fluxes of
positive and negative muons 
 in a momentum range from 0.6 to 20~GeV/$c$ 
with systematic errors less than 3~\% in both measurements.
By comparing the data collected at two different geomagnetic latitudes,
we have seen an effect of cutoff rigidity.
The dependence on the atmospheric pressure and temperature, and the solar modulation effect have been also
clearly observed.
We also clearly observed the decrease of 
charge ratio of muons at
low momentum side with at higher cutoff rigidity region.

\end{abstract}

\end{frontmatter}

\section{Introduction}
\label{}


The evidence for atmospheric neutrino oscillation has been reported from the
Super-Kamiokande collaboration by using high-statistics samples of muon
neutrino events \cite{FU98}. 
There are two major sources of systematic errors
in evaluating the neutrino flux; the flux of primary cosmic-rays and the
production cross sections of secondary mesons; pions and kaons~\cite{GHKL96,GH2002,HON95}. 
Recently,
the fluxes of primary cosmic-ray particles, mainly consisting of protons and
helium nuclei, have been measured precisely by two independent and
consistent observations~\cite{SA99,AL2000}. 
Although the details of interaction model
itself is hard to be determined, the measurement of atmospheric muons plays
crucial role in evaluating the flux of atmospheric neutrinos because muons
and muon neutrinos are produced always in pairs as decay products of mesons
and the kinematics of meson and muon decay is well known.

The muon flux at sea level has been measured by many groups. However, there
are large discrepancies among those measurements much larger than the
statistical error quoted in each publication. 
Therefore it is conceivable
that the difference comes from systematic effects such as uncertainties in
momentum determination, geometrical factor, exposure time, particle
identification, trigger efficiency and normalization procedure.

We report here precise measurements of the absolute flux of atmospheric
muons at sea level at Tsukuba ($36.2^{\circ}N$,$140.1^{\circ}E$), Japan
and Lynn Lake ($56.5^{\circ}N$,$101.0^{\circ}W$), Canada by using the BESS
spectrometer~\cite{AJ99}. The data were collected in '95 (at Tsukuba) and in '97, '98
and '99 (at Lynn Lake). The cutoff rigidities are 11.4 GV (at Tsukuba) and
0.4 GV (at Lynn Lake).

 \section{Spectrometer Setup}
 \label{}

The BESS spectrometer was designed as a high resolution
spectrometer with a large geometrical acceptance to perform 
precise measurements of primary and secondary cosmic-rays as well as a
sensitive search for rare exotic particles of primary origin\cite{OR87,AY94}.
Cross sectional views of '95 and '99 configurations are shown in 
Fig.~1.
The spectrometer configuration was updated in '97 as described below, and
was kept similar in '98 and 99 except for shower counters installed in '99.

The thin superconducting coil \cite{YA88}
 (4.70 g/cm$^2$ thick including the cryostat)
 produces a uniform axial magnetic field of 1 Tesla.
 A jet-type drift chamber (JET), inner drift chambers (IDCs) and outer
 drift chambers (ODCs) are located inside and outside the coil.
These chambers are operated with a slow gas (CO$_2$ 90~\%, Ar 10~\%).
Tracking signal from the drift chambers are read out by flash ADCs.
The $r\phi$-tracking 
is performed
 by fitting up to 28 hit-points, each with a spatial resolution of 200 $\mu$m.
Tracking in the $z$-coordinate is made by fitting points in IDC
measured with vernier pads with an accuracy of 470 $\mu$m
 and points in the JET chamber measured using charge-division with a
 spatial resolution of 20~mm. 
By using these data, we performed the continuous and redundant
3-dimensional track information. 
In order to get momentum of particle, we used 28 hit-points of
the JET
chamber and IDCs in the magnetic field. 
The ODCs provide extra hit positions outside the magnet and is used to
calibrate the JET chamber and IDCs.
In addition, all drift chambers have capabilities to distinguish the multi-hit.
This feature enables us to recognize
 multi-track events, thus we could see the tracks having interactions and
 scatterings.
The time-of-flight (TOF) scintillator hodoscopes measured the velocity of
 particles with a time resolution of 110~ps in '95. 
The acrylic \v{C}erenkov shower counter consists of acrylic and lead plate (12~mm).
These counters are placed outside the lower TOF counter.
The acrylic \v{C}erenkov shower counters, used to separate electron and
 muons, were installed only 
for the ground observation.
The total material thickness from outside the pressure vessel, passing
 through superconducting magnet coil, 
 inside the JET chamber was 9.03 g/cm$^2$.

Since '97 experiment, we installed a newly developed threshold-type \v{C}erenkov counter with 
silica-aerogel radiator, after removing the ODCs~\cite{agel}. 
The resolution of TOF was
improved to 75~ps by using new photomultipliers (PMTs) with a larger diameter for better light
collection~\cite{tof97}.  
In '99 experiment, we installed a part of the shower counter just below
the 
superconducting magnet.

 \section{Data Samples}
 \label{}

 The '95 ``ground'' experiment was carried out at 
KEK, Tsukuba ($36.2^{\circ}N$,$140.1^{\circ}E$), Japan, 
from December 23 to 28.
KEK is located at 30~m above sea level.
 The vertical cutoff rigidity is 
11.4 GV~\cite{SheaSmart}($\lambda=26.6^{\circ}N$ at geomagnetic latitude~\cite{GEOMAG}).
 The mean atmospheric pressure in this experiment was 1010~hPa (1030~g/cm$^{2}$).
The scientific data were taken for a live time period of 
291,430 sec
 and 
 9,148,104 events were recorded on magnetic tapes.
The '97, '98 and '99 ground experiments were carried out 
in Lynn 
Lake ($56.5^{\circ}N$, $101.0^{\circ}W$), Canada, on
July 22, 
 August 16 and July 26, respectively.
The experimental site in Lynn Lake is located at 360~m above sea level.
 The vertical cutoff rigidity is 
0.4 GV~\cite{SheaSmart}($\lambda=65.5^{\circ}N$ at geomagnetic latitude~\cite{GEOMAG}).
 The mean atmospheric pressures in Lynn Lake experiments in '97, '98 and
 '99 were 980.6~hPa (1000~g/cm$^{2}$), 990.5~hPa (1010~g/cm$^{2}$) and
 964.9~hPa (983.9~g/cm$^{2}$), respectively.
 The total scientific data were obtained for a period of 21,304~sec (7,011~sec, 3,949~sec and 10,344~sec) of live time 
and 
 242,934, 137,629 and 354,869 events were recorded on the magnetic tapes, respectively.

The trigger was provided
 by a coincidence between the top and the bottom scintillators of TOF counters.
All triggered events were gathered in the magnetic tapes.
The core information (momentum, T.O.F., etc.) was composed and extracted
 from  the original data.
There were two kinds of efficiencies so as to gather atmospheric
 cosmic-ray data
; trigger efficiency
 ($\varepsilon_{trigger}$), track reconstruction efficiency
 ($\varepsilon_{reconstruction}$).

 \section{Data Analysis}
 \label{}

At first, the following off-line selections were applied for the recorded
events.

(i)~One or two counters are hit in each layer of the TOF hodoscope
and only one track should be found in the JET chamber.

(ii)~Track should be 
fully contained in the fiducial region, namely the number of hits
in the JET chamber expected from the trajectory 
should be 24 and the extrapolated track should cross
 the fiducial region of TOF scintillators ($|z|<43.0$ cm). 

We call an efficiency that pass through these selection by the name of
single track efficiency
($\varepsilon_{single}$). 
We used Monte Carlo calculation in order to obtain efficiency which
depend on the momentum.

Next, we selected muon tracks from tracks that pass through the above
selection.
In order to select the muon tracks, we used the time-of-flight 
and rigidity information obtained by the TOF scintillation counters
and drift chambers, respectively as
shown in Fig.~\ref{fig:betamom}.
We selected the muon tracks using ''muon $\beta^{-1}$-band cut'' which are
defined by :
 \[ \frac{1}{\beta} \> = \> \sqrt{\left( \frac{m}{R} \right)^{2} + 1 }\; \pm
 \; 3.89 \sigma.~~~(0.01\%)\]
Here, $\beta$ is velocity of particle, $m$ is muon mass and rigidity($R$) is momentum per charge.  
We selected particles which pass through this requirement ($< | 3.89\sigma
| $),  and we call this selection efficiency  muon selection efficiency
($\varepsilon_{\mu -select}$).
From the plots, 
protons, electrons and
positrons are major sources of background events that contaminates the muon bands.

The rejection of electrons and positrons was
performed by utilizing the Acrylic \v{C}erenkov shower counter.
Proton events could be eliminated in a rigidity range below 1.4~GV by muon $\beta^{-1}$-band cut.
Above this rigidity, we used other experimental data of proton flux
\cite{brooke64,diggory74,golden95} to reduce 
the contamination of
protons into muon $\beta^{-1}$-band.
A contamination of protons in the muon $\beta^{-1}$-band was estimated
to be 2.0~\% at
1.4~GV and decreased rapidly with rigidity.
According to the work of R.~L.~Golden {\it et al.} \cite{golden95}, the
proton flux at sea level 
follows the power spectrum with an index of about $-3.0$ from 2 to 20~GV, steeper
than the index of muon flux on the ground level.
The protons were subtracted from observed muon $\beta^{-1}$-band using the
result of R.~L.~Golden {\it et al.} normalized to number of protons below
1.4~GV obtained in this experiment. 
A contamination of electrons and positrons in the muon $\beta^{-1}$-band cut
was estimated by using Acrylic \v{C}erenkov shower counter.
The electrons interacts with lead plates (12~mm) and it generates shower
in acrylic plates.
Therefore \v{C}erenkov light yielded by electrons are distinguished from
that of muons~\cite{sanuki-cerenkov}.
A contamination was estimated to be about 2~\% at 0.5 GV and decreased drastically with
rigidity, because the electron flux had steeper index than
that of the the muon index.
We subtracted electrons and positrons from muon $\beta^{-1}$-band cut 
by using the estimation obtained by analysing the Acrylic \v{C}erenkov shower counter.
The systematic errors of these subtraction was 1~\% for protons at
1.4~GeV/$c$ and 0.5~\%
for electrons and positrons at 0.6~GeV/$c$.
The systematic errors decreased drastically as rigidity increases and it
was negligible at 20~GeV/$c$.
We used muon events that pass through
only these selections to obtain muon energy spectrum.
As we shall see later,
we had about 98.9~\% efficiency to take the muon
events.

Based on these muon events, we obtained the muon rigidity spectrum at
the top of instrument (TOI) in the following way:
The TOI energy of each event was calculated by tracing back the
particle through the spectrometer material and correcting energy loss by using 
GEANT 3.21.
The corrections were usually small, 
 about 10 MeV for a 1 GeV event.

Among the factors necessary to obtain the flux,
 the geometrical acceptance can be calculated reliably
by Monte Carlo (M.C.) methods
 due to the simple geometry and the uniform magnetic field of the BESS spectrometer.
The geometrical acceptance for the vertical muons ($\cos \theta \ge 0.98$) taken 
in Tsukuba ('95)
 was about 0.03 m$^{2}$sr above 2~GeV/$c$ and decreased gradually at lower 
 momentum.
Because east and west effect is not important at high
 latitude, we analyze the data taken at Lynn Lake ('97, '98 and '99), in a
 range of  $\cos \theta \ge 0.90$ (0.09~m$^{2}$Sr).
The systematic error caused by the east-west effect in Tsukuba was estimated
to be 1.0~\% by comparison with the experimental data and the isotropic
M.C. calculation, and to be negligible in Lynn Lake.
The mean value of zenith angle distribution of muon flux was 
$\cos \theta = 0.990$ for Tsukuba data ($\cos \theta \ge 0.98$), 
and $\cos \theta = 0.955$ for Lynn Lake data ($\cos \theta \ge 0.90$).
We estimated that the total systematic error of the geometrical 
acceptance was 0.4~\%.
The geometrical acceptance can be calculated reliably
 both by an analytical method and by Monte Carlo methods at simple
 geometries, such as circle, quadrangle, etc.
 The difference of the results obtained by both
 methods was negligible (less than 0.2\%).
A systematic error of geometrical acceptance 
due to imperfect alignment was dominant.
The livetime fraction of exposure time was 94.9~\% ('95) and 99.3~\% ('97,
'98 and '99); the error due to this factor was negligibly small.

In summary, the efficiencies used in deriving the muon flux were 
trigger efficiency ($\varepsilon_{trigger}$), track reconstruction
efficiency ($\varepsilon_{reconstruction}$), 
single track efficiency ($\varepsilon_{single}$) and muon selection efficiency
($\varepsilon_{\mu -select}$).
The trigger was provided
 by a coincidence between the top and the bottom scintillators,
 with the threshold set at 1/3 of the pulse height
 from vertically incident minimum ionizing particles.
$\varepsilon_{trigger}$ was obtained from 
pulse height distribution of the TOF counter.
The efficiency for the trigger ($\varepsilon_{trigger}$) was estimated
to be 99.95~\%.
All triggered events were recorded in magnetic tape, thereafter data
summary tape (DST) was constructed by using the calibration data base.
The DST contains information of the track (momentum, track length, etc.),
therefore only reconstructed events were filled in DST and we analyzed
the muon flux by using DST.
In order to estimate 
 $\varepsilon_{reconstruction}$, we made off-line scanning (eye
 scanning) for about 1000
 tracks by using data made before DST and 
 $\varepsilon_{reconstruction}$ was found to be 99.5~\%.
 The single track selection efficiency
 ($\varepsilon_{single}$) 
was obtained from the M.C. simulation
and the systematic error was estimated 
 by examining agreements between
 observed and simulated distributions of the values
 used in the single track selection.
$\varepsilon_{single}$ was found to be 99.5~\%.
The M.C. data agreed with the real data within 1.5~\% in total.
 In order to select the muon events, we utilized the $\beta$ band cut that has 
 a width of 3.89~$\sigma$, thus the muon selection efficiency
 ($\varepsilon_{\mu -select}$) was 99.99~\%.
From the efficiencies mentioned above, the total efficiency was found to be 98.9~\%.

In order to eliminate possible influence of the momentum resolution to the muon
flux,
we used momentum up to 20~GeV/$c$.
The momentum resolution of BESS spectrometer was $\Delta P/P = 0.005P$ 
(M.D.M.$ = 200$~GeV/$c$). 
Therefore the errors of the muon flux was 1~\% at 20~GeV/$c$ by M.C. calculation if we assumed the
spectral index of muon flux is $-2.7$.
Our previous paper~\cite{SA99} discussed about this spectrum deformation effect of the BESS spectrometer.
As the momentum decreases, this error decreases.
Then the errors caused by this effect was negligible at 0.6~GeV/$c$.

Summation of all the estimated systematic errors
were 2.4~\% for positive muons and 2.2~\% for negative muons in Tsukuba and
2.2~\% for positive muons and 1.9~\% for negative muons in Lynn Lake.

\section{Atmospheric Effect}
 \label{}

Variations in cosmic-ray flux  by the change of the atmospheric
 conditions
 is called
"atmospheric effect".
It has been known that there are two main sources of this
effect~\cite{sagisaka86}
due to variations of the
atmospheric pressure and temperature. 
Denoting integral flux of the muons at depth $x_0$~(g/cm$^2$) as $I(
\overline{E_{0}},x_{0}, \theta )$, and the changes of atmospheric pressure
and temperature as $\delta P$~(mb) and $\delta T(x)$ at $x$~(g/cm$^2$) $(x < x_0)$, we have a relation
of 
\[ \delta I( \overline{E_{0}},x_{0}, \theta ) / I(
\overline{E_{0}},x_{0}, \theta ) = -\beta ( \overline{E_{0}},x_{0},
\theta ) \delta P + \int_{0}^{x_{0}} \alpha ( x, \overline{E_{0}} , x_{0} ,
\theta ) \delta T(x)dx , \]
Here,  $E_0$ is the total energy of muons at $x_0$, $\overline{E_{0}}$ is 
the threshold energy, and $\beta ( \overline{E_{0}},x_{0},\theta )$
and  $\alpha ( x, \overline{E_{0}} , x_{0} ,\theta )$ are so called ``barometric coefficient'' and ``partial temperature coefficient'',
respectively. 

In order to get the barometric coefficient, we used two sets of '95 experimental
data 
taken at different atmospheric pressures with a deviation of 25~hPa.
Fig.~\ref{fig:baro} shows the barometric coefficient for the integral
 muon flux.
The barometic effect has a negative correlation, and then flux decreases if
the atmospheric pressure increases.
A specific negative correlation due to the 
increases of the $\mu - e$ decay is
expected 
dominant below 2~GeV/$c$ and another specific negative effect due to the absorption by the
ionization loss becomes dominant above 2~GeV/$c$.
The observed coefficient seemed to be consistent with calculated values
as shown in Fig.~\ref{fig:baro}.
The effect at the 25~hPa pressure-difference on the muon flux amounts to be
2.5~\% below 1~GeV/$c$ and less than 1~\% above 5~GeV/$c$.

The temperature effect was calculated by using
a 
temperature coefficient reported by S.~Sagisaka~\cite{sagisaka86}, and observed
variations in '95 experimental data are shown in
Fig.~\ref{fig:temp}.
We used high altitude temperature data observed by 
using a radio sonde data taken at 
Tateno
Meteorological Observatory
($36.1^{\circ}N$, $140.1^{\circ}E$, 10~km south of KEK)~\cite{ADJ}. 
In order to analyse the temperature effect, we used two data sets 
which were taken at different temperature 
at the '95 experiment.
The observed variation seemed to be consistent with the calculated variation.
The variation of muon flux due to the temperature effect in the period 
of this experiment was less than 1~\%.

 \section{Solar Modulation}
 \label{}

Fig.~\ref{fig:aho3y} shows annual variation of the muon flux in Lynn Lake.
In order to distinguish small difference of each variation, the '97 and
'98 muon fluxes
are divided by the '99 muon flux, and the flux of each year were combined 
in a wide momentum region to reduce the statistical errors.
It is clearly shown that the '99 flux is lower than other fluxes.
The '99 experiment was performed at the lowest ambient pressure among
other three measurements.
Since the barometric effect has the negative correlation, 
the reduction of the flux in '99 can not be explained by
the barometric effect.
We need to takes into account an effect of  the solar modulation.
The solar activity varies globally with the 11 year solar cycle and
the solar minimum was '96 - '97 and the solar maximum 
happened between '00 and '01
according to observations of sunspot numbers~\cite{nasasp}.
However, this effect appears about one year later in neutron
monitor data~\cite{climax}.
Not only the muon flux, but also muon charge ratios ($\mu^+ / \mu^-$)
decrease below 3.5~GV if the low
energy primary proton 
flux decreases by the solar modulation.
These charge ratios of '97, '98 and '99 experiments in this energy region (0.58 -
3.44 GeV/$c$) were $1.258
\pm 0.017$, $1.235 \pm 0.022$ and $1.218 \pm 0.014$. These
decreases were consistent to decreases of the muon flux.
The decrease of charge ratios and muon fluxes are caused by decreasing of
primary proton flux, therefore the effect of solar modulation were
observed. 
The BESS spectrometer observed variation of the primary proton flux due to
the solar
modulation effect at an altitude of 37~km (launched from Lynn Lake) from '97
to '00~\cite{sanukid}. 
These variation shows about 20~\% decrease at
2~GeV/$c$ from '97 to '99 and the difference of these fluxes becomes
much smaller at
higher momentum. 
The mean energy of the primary proton responsible to muons of
0.5~GeV/$c$ at sea level is about 20 - 30~GeV, and then the degree of muon
flux is much smaller than that of primary proton flux at the same energy.

These difference of muon flux were 3~\% around 1~GeV/$c$.
These differences were within statistic and systematic errors with small
bins and 
it was important to obtain spectral shape with small statistic errors.
Therefore '97, '98 and '99 experimental data sets were combined to
obtain the muon flux in Lynn Lake.

 \section{Results}
 \label{}

Fig.~\ref{fig:ahopm} shows the resultant positive and negative muon
fluxes, and Table.~\ref{tab:muonflux95} and Table.~\ref{tab:muonflux99}
summarize those data with systematic and statistic errors.
 We have observed the vertical fluxes of the 
positive and negative muons
 in a momentum range from 0.6 to 20~GeV/$c$ 
 with an estimated systematic 
error of 2.4~\% for the positive muons and 2.2~\% for the negative muons in
Tsukuba, and 2.2~\% for the positive muons and 1.9~\% for the negative muons in
Lynn Lake. 
The cutoff rigidity at Tsukuba is much higher than Lynn Lake.
By comparing the data collected at two different geomagnetic latitudes,
we have observed an effect of cutoff rigidity.

Fig.~\ref{fig:ahof} shows the total (positive and negative) differential
 muon spectra at Tsukuba and Lynn Lake, 
 together with previous measurements~\cite{HW62,GD79,AB75,NS72,AC71,BC71,R84,TKO98,DMP93,KBA99}.
Our data on the muon fluxes those which were multiplied by $p^2$ at sea level are shown in
 Fig.~\ref{fig:ahof2}.
From these figures, it is  clearly seen that the muon flux measured in
 Tsukuba and in Lynn Lake were different in lower momentum ranged below 3.5~GeV/$c$, but 
were in good agreement in higher momentum beyond 3.5~GeV/$c$.
This is because the cutoff rigidity for primary cosmic-rays does not affect
 in higher momentum.

Fig.~\ref{fig:ahor} shows ratios of positive and negative muons  together with
the previous measurements~\cite{KBA99,AHR71,BHT75,NS722,R842}, and the results
are summarized in Table.~\ref{tab:muonratio95} and
in Table.~\ref{tab:muonratio99}. 
It was seen
that the charge ratio obtained in Tsukuba decreased below 3.5~GeV/$c$
while the charge ratio obtained in Lynn Lake remained almost constant
value even in this energy range.
This difference comes from the influence of the geomagnetic cutoff
rigidity.
Because the low momentum muons must be generated by the higher
momentum protons at Tsukuba.
The muon charge ratio observed in Tsukuba had to include systematic
errors of 
proton subtraction and 
east-west effect.
On the other hand, the muon charge ratio observed in Lynn Lake had systematic errors due to only
proton subtraction.
The systematic error due to proton contamination was less 
than 2~\%, therefore the muon
charge ratio observed in Lynn Lake had very small systematic errors.

 \section{Discussion}
 \label{}

The obtained momentum spectrum appeared to be good agreement with
recent CAPRICE~94~\cite{KBA99} data using the instruments of magnetic spectrometer.
These data agreed well within the systematic and statistic errors.
But the results of previous experiments were about 20~\% larger than these recent
experimental data.
Most of these previous experiments needed normalization point in order to
determine absolute muon fluxes.
Since the spectrum shape is similar enough among those experiments,
systematic errors of the absolute fluxes are supposed to be the main cause of these
difference of absolute fluxes. 
Therefore we performed our observations with a great care to evaluate
the efficiencies to detect the muon tracks.
We then have muon fluxes with much smaller systematic errors.

The atmospheric effect was clearly observed.
Our result agreed well with the expectation of the analytical calculation\cite{sagisaka86}.
The barometric coefficient had about $-0.1$ \%/hPa at 1~GeV/$c$.
A temperature effect had less influence on
the muon flux in comparison with the barometric effect.
Our
observation of the temperature effect 
can also interpreted quantitatively with the expectation of 
an analytical calculation\cite{sagisaka86}.
Therefore for a precise calculation of the atmospheric neutrino flux
precisely, 
we could include these effect in an analytical way.

The solar modulation effects to the muon flux at the ground level was
clearly observed in our experiments. 
Not only decreasing of the total flux, but also
decreasing of the muon charge ratio has been observed.
Decreasing of the muon flux should be due to decreasing of primary
proton flux according to a temporal variation of the solar modulation.
In the atmospheric neutrino flux calculation, it may be important to
consider even small changes of the muon fluxes caused by the solar modulation.

 \section{Conclusion}
 \label{}

The vertical absolute fluxes of atmospheric muons have been precisely
measured with systematic errors of 2.4~\% or smaller.
We observed the geomagnetic effect by comparing the muon fluxes observed
at Tsukuba, Japan and Lynn Lake, Canada.
Muon charge ratios obtained at these two sites also showed the geomagnetic effect.
 The precise measurement of the muon flux at sea level is very 
important to understand cosmic-ray interactions inside the
 atmosphere and to decide fundamental parameters to study atmospheric neutrino oscillation.

\ack

BESS experiment has been supported by Grants-in-Aid from 
Ministry of Education, Culture, Sports,Science and Technology, (MEXT) and
Heiwa Nakajima Foundation in Japan and by NASA in the U.S.A.
The analysis was performed with the computing facilities at ICEPP, 
University of Tokyo.
The data of the CLIMAX neutron monitor were provided by Space Physical
Data System of University of Chicago, supported by National Science
Foundation Grant ATM-9912341.


\clearpage

\begin{figure}
 \label{fig:detectors95-99}
 \begin{minipage}{.45\textwidth}
  \begin{center}
    \vspace*{0.3cm}
   \includegraphics[width=5cm]{bess95.epsi}
    \vspace*{0.6cm}
  \\ BESS '95.
  \end{center}
 \end{minipage}
 \begin{minipage}{.45\textwidth}
  \begin{center}
   \includegraphics[width=6.3cm]{besscross-shw.epsi}
  \\ BESS '99.
  \end{center}
 \end{minipage}
 \caption{Cross-sectional view of the BESS '95 and '99 spectrometers.}
\end{figure}

\clearpage

\begin{figure}[t]
\begin{center}
\includegraphics[width=12cm]{betamom.epsi}
\end{center}
\caption{Scatter plots of $\beta^{-1}$ vs. rigidity for positively and 
negatively charged particles. ('95).}
\label{fig:betamom}
\end{figure}

\clearpage

\begin{figure}[t]
\begin{center}
\includegraphics[width=12cm]{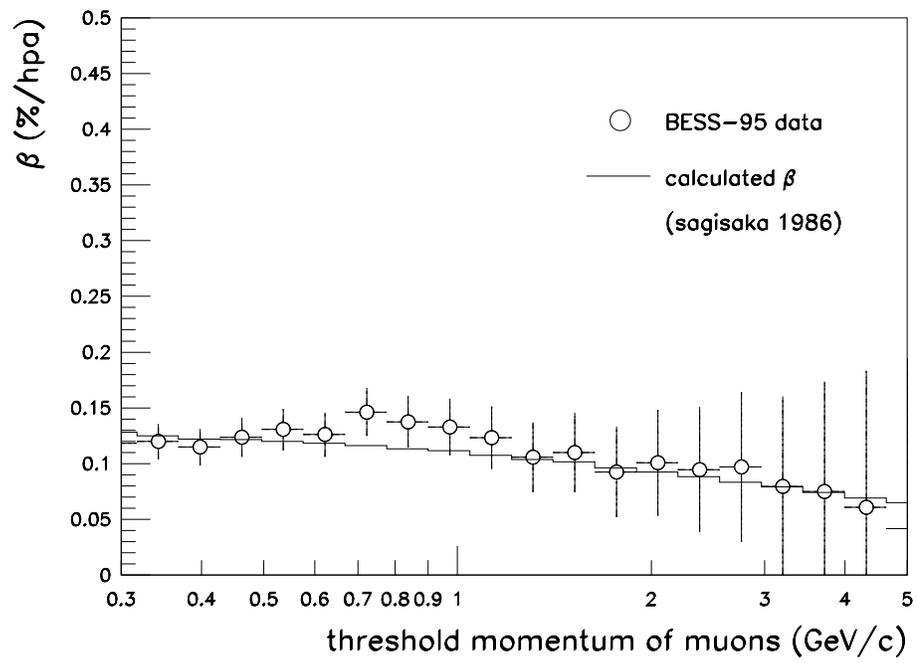}
\end{center}
\caption{Barometric coefficient.
  }
\label{fig:baro}
\end{figure}

\clearpage

\begin{figure}[t]
\begin{center}
\includegraphics[bb=28 403 513 760,clip,width=12cm]{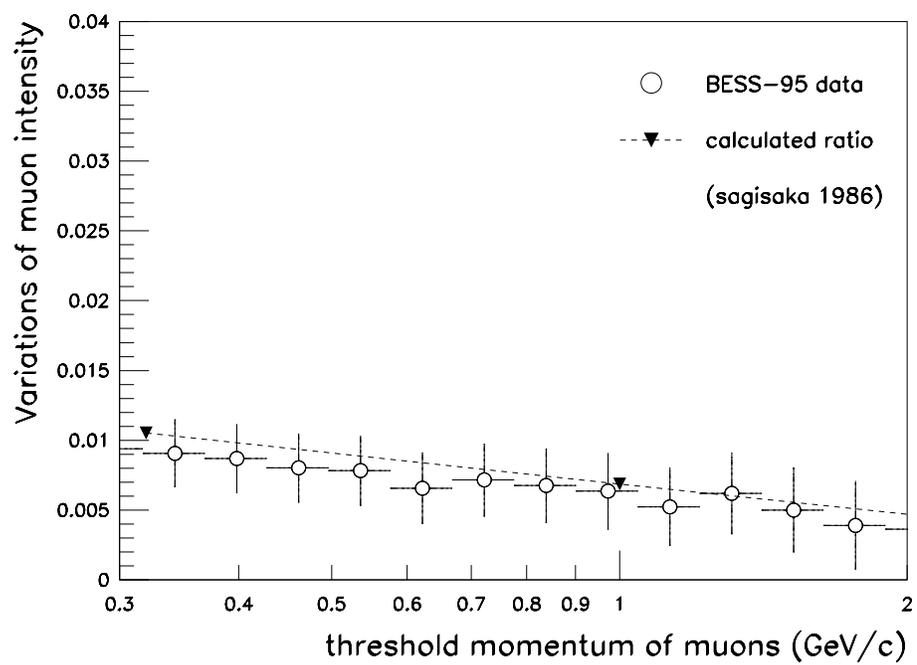}
\end{center}
\caption{Flux ratio due to temperature effect.
  }
\label{fig:temp}
\end{figure}

\clearpage

\begin{figure}[t]
\begin{center}
\includegraphics[width=12cm]{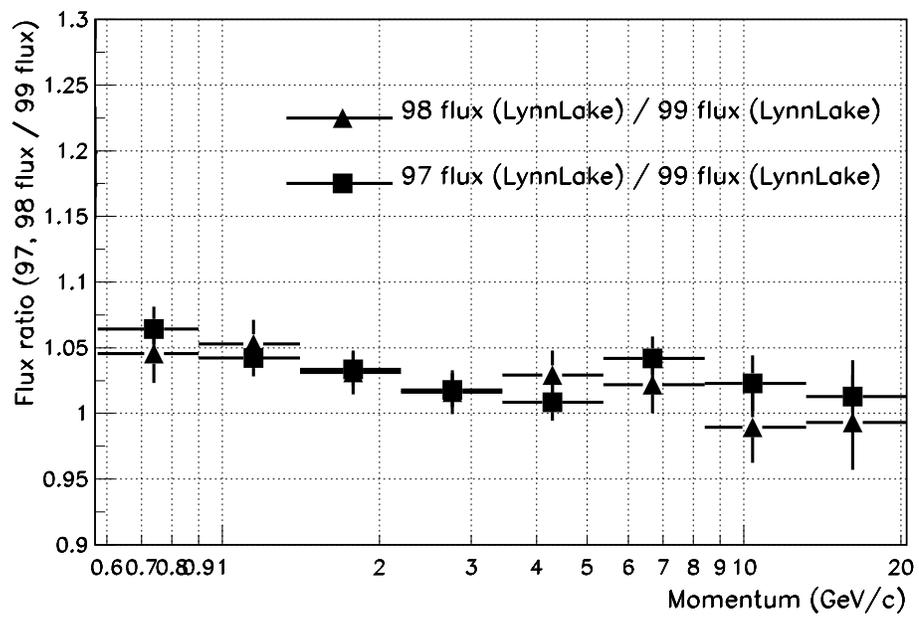}
\end{center}
\caption{Annual variation of the muon flux.
  }
\label{fig:aho3y}
\end{figure}

\clearpage

\begin{figure}[t]
\begin{center}
\includegraphics[width=10cm]{ahopm.epsi}
\\
Lynn Lake ('97, '98 and '99). \\
Tsukuba ('95).
\end{center}
\caption{BESS results for vertical differential momentum spectra of the 
 positive and negative muons at sea level .
  }
\label{fig:ahopm}
\end{figure}

\clearpage

\begin{figure}[t]
\begin{center}
\includegraphics[width=10cm]{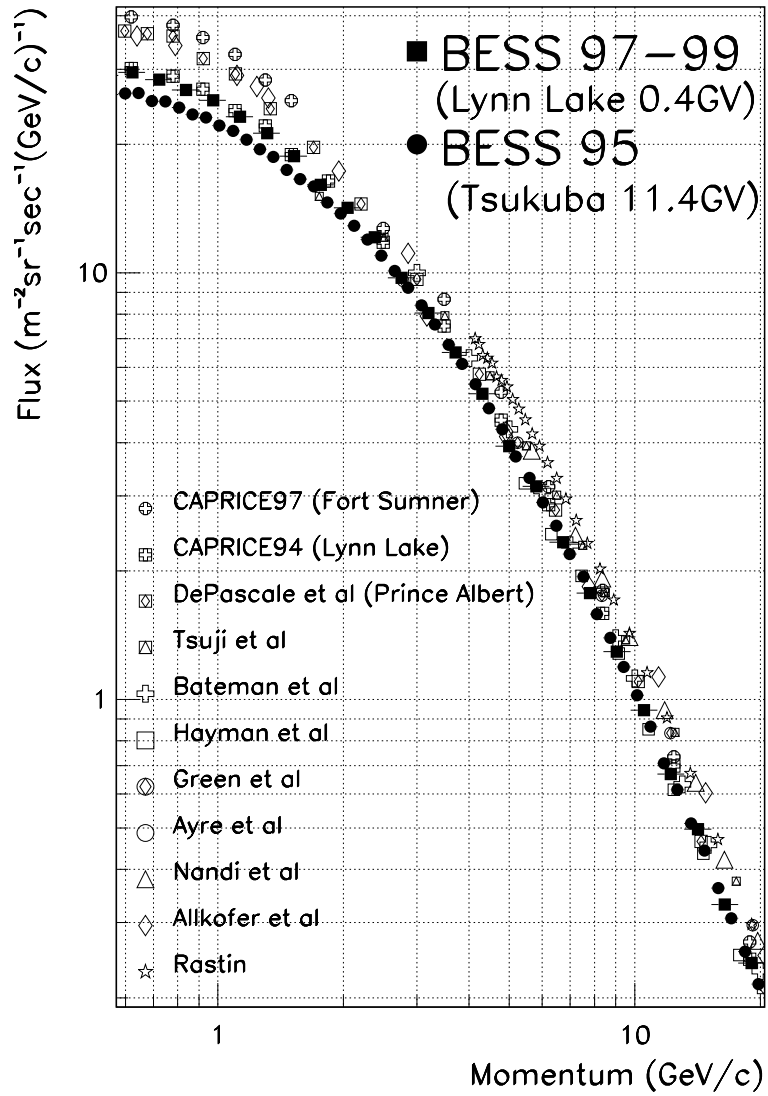}
\end{center}
\caption{BESS results of vertical differential momentum spectrum of muons at sea level 
  together with previous data.
  }
\label{fig:ahof}
\end{figure}

\clearpage

\begin{figure}[t]
\begin{center}
\includegraphics[width=13cm]{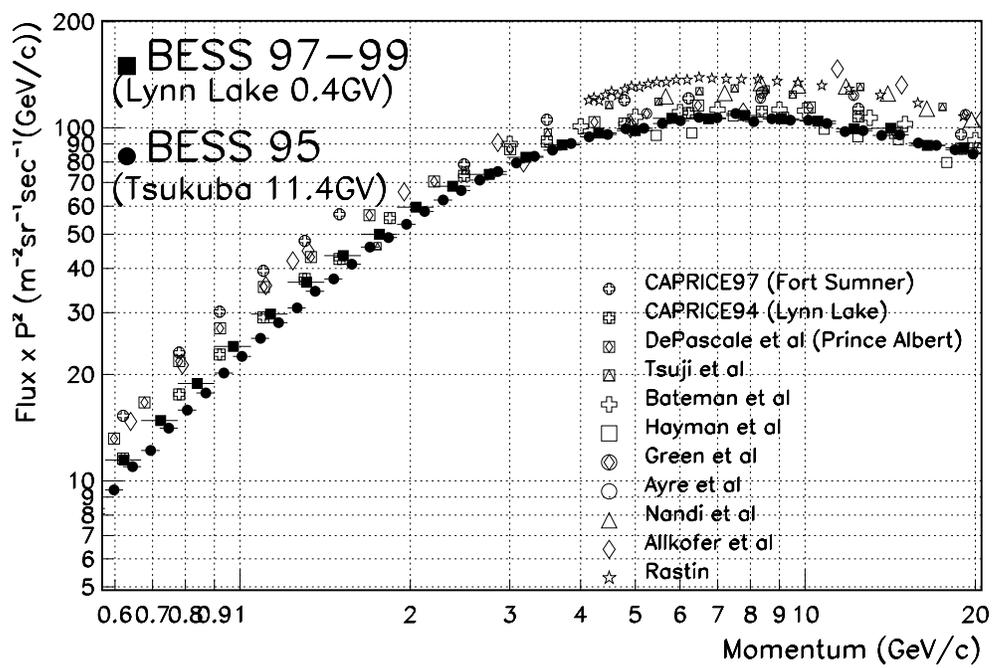}
\end{center}
\caption{The muon fluxes measured by BESS}
\label{fig:ahof2}
\end{figure}

\clearpage

\begin{figure}[t]
\begin{center}
\includegraphics[width=13cm]{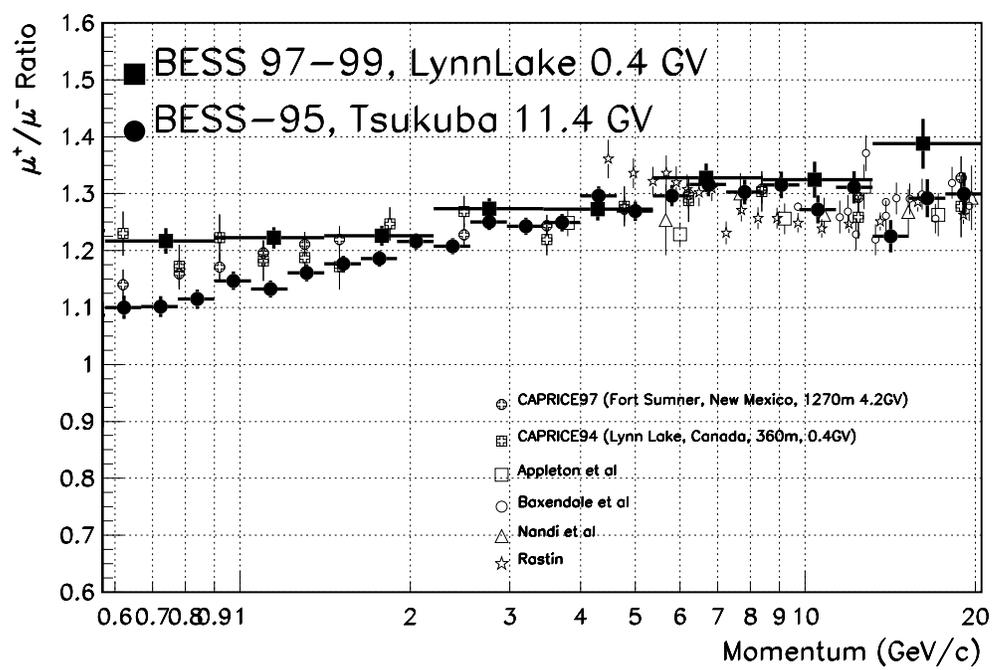}
\end{center}
\caption{BESS results of muon charge ratio at sea level 
  together with previous data.
  }
\label{fig:ahor}
\end{figure}

\renewcommand{\baselinestretch}{1.0}

\begin{table}[hb]
  \caption
  [Positive Muon Flux (Tsukuba '95).]
  {Positive Muon Flux (Tsukuba '95).}
  \label{tab:muonflux95}
  \begin{center}
    \begin{tabular}{lcccc}
      \hline
      \hline
  \multicolumn{5}{c}{Tsukuba, Japan} \\
      \hline
Momentum & Mean     & $\mu^+$            &   &     \\
 Range & Momentum & Differential flux & Statistical Error & Systematic Error  \\
  (GeV/$c$) &  (GeV/$c$) & (m$^{-2}$sr$^{-1}$sec$^{-1}$(GeV/c)$^{-1}$) &     &    \\
      \hline                                                            

0.576-0.621 & 0.598 & 1.386e+01 & 2.6e-01 & 2.7e-01 \\ 
0.621-0.669 & 0.645 & 1.375e+01 & 2.4e-01 & 2.6e-01 \\ 
0.669-0.720 & 0.695 & 1.310e+01 & 2.2e-01 & 2.5e-01 \\ 
0.720-0.776 & 0.748 & 1.332e+01 & 2.1e-01 & 2.5e-01 \\ 
0.776-0.836 & 0.806 & 1.287e+01 & 2.0e-01 & 2.5e-01 \\ 
0.836-0.901 & 0.868 & 1.236e+01 & 1.8e-01 & 2.4e-01 \\ 
0.901-0.970 & 0.936 & 1.223e+01 & 1.7e-01 & 2.3e-01 \\ 
0.970-1.045 & 1.008 & 1.192e+01 & 1.6e-01 & 2.3e-01 \\ 
1.045-1.126 & 1.086 & 1.137e+01 & 1.5e-01 & 2.2e-01 \\ 
1.126-1.213 & 1.170 & 1.091e+01 & 1.4e-01 & 2.1e-01 \\ 
1.213-1.307 & 1.260 & 1.055e+01 & 1.3e-01 & 2.0e-01 \\ 
1.307-1.408 & 1.357 & 9.951e+00 & 1.2e-01 & 1.9e-01 \\ 
1.408-1.517 & 1.463 & 9.390e+00 & 1.1e-01 & 2.0e-01 \\ 
1.517-1.634 & 1.575 & 8.989e+00 & 1.1e-01 & 1.9e-01 \\ 
1.634-1.760 & 1.697 & 8.613e+00 & 1.0e-01 & 1.8e-01 \\ 
1.760-1.896 & 1.828 & 7.962e+00 & 9.1e-02 & 1.7e-01 \\ 
1.896-2.043 & 1.969 & 7.519e+00 & 8.6e-02 & 1.6e-01 \\ 
2.043-2.201 & 2.121 & 7.094e+00 & 8.0e-02 & 1.5e-01 \\ 
2.201-2.371 & 2.285 & 6.543e+00 & 7.3e-02 & 1.4e-01 \\ 
2.371-2.555 & 2.462 & 6.000e+00 & 6.8e-02 & 1.2e-01 \\ 
2.555-2.752 & 2.653 & 5.596e+00 & 6.3e-02 & 1.2e-01 \\ 
2.752-2.965 & 2.857 & 5.139e+00 & 5.8e-02 & 1.1e-01 \\ 
2.965-3.194 & 3.078 & 4.622e+00 & 5.3e-02 & 9.5e-02 \\ 
3.194-3.441 & 3.315 & 4.212e+00 & 4.9e-02 & 8.6e-02 \\ 
3.441-3.707 & 3.573 & 3.742e+00 & 4.4e-02 & 7.6e-02 \\ 
3.707-3.993 & 3.847 & 3.417e+00 & 4.0e-02 & 6.9e-02 \\ 
3.993-4.302 & 4.145 & 3.089e+00 & 3.7e-02 & 6.3e-02 \\ 

\hline
    \end{tabular}
  \end{center}
\end{table}

\addtocounter{table}{-1}

\begin{table}[hb]
  \caption
  [Positive Muon Flux (Tsukuba '95).]
  {Positive Muon Flux (Tsukuba '95).}
  \label{tab:muonflux95}
  \begin{center}
    \begin{tabular}{lcccc}
      \hline
      \hline
  \multicolumn{5}{c}{Tsukuba, Japan} \\
      \hline
Momentum & Mean     & $\mu^+$            &   &     \\
 Range & Momentum & Differential flux & Statistical Error & Systematic Error  \\
  (GeV/$c$) &  (GeV/$c$) & (m$^{-2}$sr$^{-1}$sec$^{-1}$(GeV/c)$^{-1}$) &     &    \\
      \hline

4.302-4.635 & 4.465 & 2.719e+00 & 3.4e-02 & 5.5e-02 \\ 
4.635-4.993 & 4.809 & 2.419e+00 & 3.0e-02 & 4.9e-02 \\ 
4.993-5.379 & 5.182 & 2.060e+00 & 2.7e-02 & 4.2e-02 \\ 
5.379-5.795 & 5.583 & 1.873e+00 & 2.5e-02 & 3.8e-02 \\ 
5.795-6.243 & 6.016 & 1.628e+00 & 2.2e-02 & 3.3e-02 \\ 
6.243-6.726 & 6.478 & 1.448e+00 & 2.0e-02 & 2.9e-02 \\ 
6.726-7.246 & 6.983 & 1.248e+00 & 1.8e-02 & 2.5e-02 \\ 
7.246-7.806 & 7.519 & 1.101e+00 & 1.6e-02 & 2.2e-02 \\ 
7.806-8.409 & 8.099 & 8.934e-01 & 1.4e-02 & 1.8e-02 \\ 
8.409-9.059 & 8.728 & 7.962e-01 & 1.3e-02 & 1.6e-02 \\ 
9.059-9.760 & 9.399 & 6.731e-01 & 1.2e-02 & 1.4e-02 \\ 
9.760-10.514 & 10.126 & 5.634e-01 & 1.0e-02 & 1.1e-02 \\ 
10.514-11.327 & 10.907 & 4.923e-01 & 9.2e-03 & 1.0e-02 \\ 
11.327-12.203 & 11.754 & 3.982e-01 & 7.9e-03 & 8.1e-03 \\ 
12.203-13.146 & 12.652 & 3.521e-01 & 7.2e-03 & 7.2e-03 \\ 
13.146-14.163 & 13.649 & 2.790e-01 & 6.1e-03 & 5.7e-03 \\ 
14.163-15.258 & 14.693 & 2.465e-01 & 5.5e-03 & 5.1e-03 \\ 
15.258-16.437 & 15.826 & 2.016e-01 & 4.8e-03 & 4.2e-03 \\ 
16.437-17.708 & 17.054 & 1.752e-01 & 4.3e-03 & 3.6e-03 \\ 
17.708-19.077 & 18.378 & 1.440e-01 & 3.8e-03 & 3.0e-03 \\ 
19.077-20.552 & 19.791 & 1.227e-01 & 3.3e-03 & 2.6e-03 \\ 

\hline
    \end{tabular}
  \end{center}
\end{table}

\addtocounter{table}{-1}

\begin{table}[hb]
  \caption
  [Negative Muon Flux (Tsukuba '95).]
  {Negative Muon Flux (Tsukuba '95).}
  \label{tab:muonflux95}
  \begin{center}
    \begin{tabular}{lcccc}
      \hline
      \hline
  \multicolumn{5}{c}{Tsukuba, Japan} \\
      \hline
Momentum & Mean     & $\mu^-$            &   &     \\
 Range & Momentum & Differential flux & Statistical Error & Systematic Error  \\
  (GeV/$c$) &  (GeV/$c$) & (m$^{-2}$sr$^{-1}$sec$^{-1}$(GeV/c)$^{-1}$) &     &    \\
      \hline                                                            

0.576-0.621 & 0.598 & 1.245e+01 & 2.5e-01 & 2.4e-01 \\ 
0.621-0.669 & 0.645 & 1.262e+01 & 2.3e-01 & 2.4e-01 \\ 
0.669-0.720 & 0.695 & 1.215e+01 & 2.1e-01 & 2.3e-01 \\ 
0.720-0.776 & 0.748 & 1.186e+01 & 2.0e-01 & 2.3e-01 \\ 
0.776-0.836 & 0.806 & 1.149e+01 & 1.9e-01 & 2.2e-01 \\ 
0.836-0.901 & 0.868 & 1.113e+01 & 1.7e-01 & 2.1e-01 \\ 
0.901-0.970 & 0.936 & 1.086e+01 & 1.6e-01 & 2.1e-01 \\ 
0.970-1.045 & 1.008 & 1.021e+01 & 1.5e-01 & 1.9e-01 \\ 
1.045-1.126 & 1.086 & 1.012e+01 & 1.4e-01 & 1.9e-01 \\ 
1.126-1.213 & 1.170 & 9.572e+00 & 1.3e-01 & 1.8e-01 \\ 
1.213-1.307 & 1.260 & 8.920e+00 & 1.2e-01 & 1.7e-01 \\ 
1.307-1.408 & 1.357 & 8.722e+00 & 1.1e-01 & 1.7e-01 \\ 
1.408-1.517 & 1.463 & 8.039e+00 & 1.0e-01 & 1.5e-01 \\ 
1.517-1.634 & 1.575 & 7.590e+00 & 9.8e-02 & 1.4e-01 \\ 
1.634-1.760 & 1.697 & 7.317e+00 & 9.2e-02 & 1.4e-01 \\ 
1.760-1.896 & 1.828 & 6.662e+00 & 8.4e-02 & 1.3e-01 \\ 
1.896-2.043 & 1.969 & 6.234e+00 & 7.9e-02 & 1.2e-01 \\ 
2.043-2.201 & 2.121 & 5.787e+00 & 7.2e-02 & 1.1e-01 \\ 
2.201-2.371 & 2.285 & 5.421e+00 & 6.7e-02 & 1.0e-01 \\ 
2.371-2.555 & 2.462 & 4.966e+00 & 6.2e-02 & 9.5e-02 \\ 
2.555-2.752 & 2.653 & 4.510e+00 & 5.7e-02 & 8.6e-02 \\ 
2.752-2.965 & 2.857 & 4.075e+00 & 5.2e-02 & 7.8e-02 \\ 
2.965-3.194 & 3.078 & 3.757e+00 & 4.8e-02 & 7.2e-02 \\ 
3.194-3.441 & 3.315 & 3.353e+00 & 4.4e-02 & 6.5e-02 \\ 
3.441-3.707 & 3.573 & 3.041e+00 & 4.0e-02 & 5.9e-02 \\ 
3.707-3.993 & 3.847 & 2.694e+00 & 3.6e-02 & 5.2e-02 \\ 
3.993-4.302 & 4.145 & 2.393e+00 & 3.3e-02 & 4.6e-02 \\ 

\hline
    \end{tabular}
  \end{center}
\end{table}

\addtocounter{table}{-1}

\begin{table}[hb]
  \caption
  [Negative Muon Flux (Tsukuba '95).]
  {Negative Muon Flux (Tsukuba '95).}
  \label{tab:muonflux95}
  \begin{center}
    \begin{tabular}{lcccc}
      \hline
      \hline
  \multicolumn{5}{c}{Tsukuba, Japan} \\
      \hline
Momentum & Mean     & $\mu^-$            &   &     \\
 Range & Momentum & Differential flux & Statistical Error & Systematic Error  \\
  (GeV/$c$) &  (GeV/$c$) & (m$^{-2}$sr$^{-1}$sec$^{-1}$(GeV/c)$^{-1}$) &     &    \\
      \hline

4.302-4.635 & 4.465 & 2.090e+00 & 2.9e-02 & 4.1e-02 \\ 
4.635-4.993 & 4.809 & 1.878e+00 & 2.7e-02 & 3.7e-02 \\ 
4.993-5.379 & 5.182 & 1.649e+00 & 2.4e-02 & 3.2e-02 \\ 
5.379-5.795 & 5.583 & 1.430e+00 & 2.2e-02 & 2.8e-02 \\ 
5.795-6.243 & 6.016 & 1.270e+00 & 2.0e-02 & 2.5e-02 \\ 
6.243-6.726 & 6.478 & 1.104e+00 & 1.8e-02 & 2.2e-02 \\ 
6.726-7.246 & 6.983 & 9.449e-01 & 1.6e-02 & 1.9e-02 \\ 
7.246-7.806 & 7.519 & 8.376e-01 & 1.4e-02 & 1.7e-02 \\ 
7.806-8.409 & 8.099 & 6.926e-01 & 1.3e-02 & 1.4e-02 \\ 
8.409-9.059 & 8.728 & 5.978e-01 & 1.1e-02 & 1.2e-02 \\ 
9.059-9.760 & 9.399 & 5.188e-01 & 1.0e-02 & 1.0e-02 \\ 
9.760-10.514 & 10.126 & 4.598e-01 & 9.2e-03 & 9.3e-03 \\ 
10.514-11.327 & 10.907 & 3.713e-01 & 7.9e-03 & 7.5e-03 \\ 
11.327-12.203 & 11.754 & 3.101e-01 & 7.0e-03 & 6.3e-03 \\ 
12.203-13.146 & 12.652 & 2.625e-01 & 6.2e-03 & 5.4e-03 \\ 
13.146-14.163 & 13.649 & 2.335e-01 & 5.6e-03 & 4.8e-03 \\ 
14.163-15.258 & 14.693 & 1.958e-01 & 5.0e-03 & 4.0e-03 \\ 
15.258-16.437 & 15.826 & 1.599e-01 & 4.3e-03 & 3.3e-03 \\ 
16.437-17.708 & 17.054 & 1.320e-01 & 3.8e-03 & 2.7e-03 \\ 
17.708-19.077 & 18.378 & 1.126e-01 & 3.4e-03 & 2.4e-03 \\ 
19.077-20.552 & 19.791 & 9.271e-02 & 2.9e-03 & 1.9e-03 \\ 

\hline
    \end{tabular}
  \end{center}
\end{table}

\begin{table}[hb]
  \caption
  [Positive Muon Flux (Lynn Lake '97,'98,'99).]
  {Positive Muon Flux (Lynn Lake '97,'98,'99).}
  \label{tab:muonflux99}
  \begin{center}
    \begin{tabular}{lcccc}
      \hline
      \hline
  \multicolumn{5}{c}{Lynn Lake, Canada} \\
      \hline
Momentum & Mean     & $\mu^+$            &   &     \\
 Range & Momentum & Differential flux & Statistical Error & Systematic Error  \\
  (GeV/$c$) &  (GeV/$c$) & (m$^{-2}$sr$^{-1}$sec$^{-1}$(GeV/c)$^{-1}$) &     &    \\
      \hline                                                            

0.576-0.669 & 0.622 & 1.620e+01 & 3.9e-01 & 2.6e-01 \\ 
0.669-0.776 & 0.723 & 1.526e+01 & 3.4e-01 & 2.5e-01 \\ 
0.776-0.901 & 0.839 & 1.494e+01 & 3.0e-01 & 2.4e-01 \\ 
0.901-1.045 & 0.973 & 1.405e+01 & 2.6e-01 & 2.3e-01 \\ 
1.045-1.213 & 1.128 & 1.280e+01 & 2.3e-01 & 2.1e-01 \\ 
1.213-1.408 & 1.309 & 1.157e+01 & 2.0e-01 & 1.9e-01 \\ 
1.408-1.634 & 1.519 & 1.047e+01 & 1.7e-01 & 2.0e-01 \\ 
1.634-1.896 & 1.763 & 8.831e+00 & 1.5e-01 & 1.6e-01 \\ 
1.896-2.201 & 2.046 & 7.744e+00 & 1.3e-01 & 1.4e-01 \\ 
2.201-2.555 & 2.373 & 6.844e+00 & 1.1e-01 & 1.2e-01 \\ 
2.555-2.965 & 2.752 & 5.354e+00 & 9.0e-02 & 9.6e-02 \\ 
2.965-3.441 & 3.194 & 4.541e+00 & 7.7e-02 & 8.1e-02 \\ 
3.441-3.993 & 3.705 & 3.640e+00 & 6.4e-02 & 6.4e-02 \\ 
3.993-4.635 & 4.299 & 2.928e+00 & 5.3e-02 & 5.2e-02 \\ 
4.635-5.379 & 4.991 & 2.190e+00 & 4.3e-02 & 3.8e-02 \\ 
5.379-6.243 & 5.795 & 1.789e+00 & 3.6e-02 & 3.1e-02 \\ 
6.243-7.246 & 6.718 & 1.340e+00 & 2.9e-02 & 2.3e-02 \\ 
7.246-8.409 & 7.790 & 1.017e+00 & 2.3e-02 & 1.8e-02 \\ 
8.409-9.760 & 9.046 & 7.332e-01 & 1.8e-02 & 1.3e-02 \\ 
9.760-11.327 & 10.504 & 5.437e-01 & 1.5e-02 & 9.6e-03 \\ 
11.327-13.146 & 12.172 & 3.784e-01 & 1.1e-02 & 6.8e-03 \\ 
13.146-15.258 & 14.144 & 2.838e-01 & 9.1e-03 & 5.1e-03 \\ 
15.258-17.708 & 16.385 & 1.924e-01 & 7.0e-03 & 3.5e-03 \\ 
17.708-20.552 & 19.078 & 1.436e-01 & 5.6e-03 & 2.6e-03 \\ 
\hline
    \end{tabular}
  \end{center}
\end{table}

\addtocounter{table}{-1}

\begin{table}[hb]
  \caption
  [Negative Muon Flux (Lynn Lake '97,'98,'99).]
  {Negative Muon Flux (Lynn Lake '97,'98,'99).}
  \label{tab:muonflux99}
  \begin{center}
    \begin{tabular}{lcccc}
      \hline
      \hline
  \multicolumn{5}{c}{Lynn Lake, Canada} \\
      \hline
Momentum & Mean     & $\mu^-$            &   &     \\
 Range & Momentum & Differential flux & Statistical Error & Systematic Error  \\
  (GeV/$c$) &  (GeV/$c$) & (m$^{-2}$sr$^{-1}$sec$^{-1}$(GeV/c)$^{-1}$) &     &    \\
      \hline                                                            

0.576-0.669 & 0.622 & 1.327e+01 & 3.4e-01 & 2.2e-01 \\ 
0.669-0.776 & 0.723 & 1.307e+01 & 3.0e-01 & 2.1e-01 \\ 
0.776-0.901 & 0.839 & 1.189e+01 & 2.6e-01 & 1.9e-01 \\ 
0.901-1.045 & 0.973 & 1.129e+01 & 2.3e-01 & 1.8e-01 \\ 
1.045-1.213 & 1.128 & 1.041e+01 & 2.1e-01 & 1.7e-01 \\ 
1.213-1.408 & 1.309 & 9.649e+00 & 1.8e-01 & 1.6e-01 \\ 
1.408-1.634 & 1.519 & 8.295e+00 & 1.5e-01 & 1.3e-01 \\ 
1.634-1.896 & 1.763 & 7.229e+00 & 1.3e-01 & 1.2e-01 \\ 
1.896-2.201 & 2.046 & 6.456e+00 & 1.2e-01 & 1.0e-01 \\ 
2.201-2.555 & 2.373 & 5.256e+00 & 9.6e-02 & 8.6e-02 \\ 
2.555-2.965 & 2.752 & 4.368e+00 & 8.1e-02 & 7.1e-02 \\ 
2.965-3.441 & 3.194 & 3.506e+00 & 6.7e-02 & 5.8e-02 \\ 
3.441-3.993 & 3.705 & 2.861e+00 & 5.7e-02 & 4.7e-02 \\ 
3.993-4.635 & 4.299 & 2.281e+00 & 4.7e-02 & 3.8e-02 \\ 
4.635-5.379 & 4.991 & 1.735e+00 & 3.8e-02 & 2.9e-02 \\ 
5.379-6.243 & 5.795 & 1.369e+00 & 3.1e-02 & 2.3e-02 \\ 
6.243-7.246 & 6.718 & 9.978e-01 & 2.5e-02 & 1.7e-02 \\ 
7.246-8.409 & 7.790 & 7.590e-01 & 2.0e-02 & 1.3e-02 \\ 
8.409-9.760 & 9.046 & 5.605e-01 & 1.6e-02 & 9.7e-03 \\ 
9.760-11.327 & 10.504 & 4.000e-01 & 1.3e-02 & 7.0e-03 \\ 
11.327-13.146 & 12.172 & 2.900e-01 & 9.9e-03 & 5.1e-03 \\ 
13.146-15.258 & 14.144 & 2.130e-01 & 7.9e-03 & 3.8e-03 \\ 
15.258-17.708 & 16.385 & 1.381e-01 & 5.9e-03 & 2.5e-03 \\ 
17.708-20.552 & 19.078 & 9.773e-02 & 4.6e-03 & 1.8e-03 \\ 

\hline
    \end{tabular}
  \end{center}
\end{table}

\begin{table}[hb]
  \caption
  [Muon Charge Ratio (Tsukuba '95).]
  {Muon Charge Ratio (Tsukuba '95).}
  \label{tab:muonratio95}
  \begin{center}
    \begin{tabular}{llccc}
      \hline
      \hline
  \multicolumn{5}{c}{Tsukuba, Japan} \\
      \hline
Momentum & Mean  &    &      &       \\
Range & Momentum & $\mu^+ / \mu^-$ Ratio & Statistical Error & Systematic Error \\
(GeV/c) &  (GeV/c) &  &   &       \\
      \hline

0.576-0.669 & 0.623 & 1.100 & 0.020 & 0.011 \\ 
0.669-0.776 & 0.723 & 1.101 & 0.018 & 0.011 \\ 
0.776-0.901 & 0.838 & 1.115 & 0.017 & 0.011 \\ 
0.901-1.045 & 0.973 & 1.147 & 0.016 & 0.011 \\ 
1.045-1.213 & 1.129 & 1.132 & 0.015 & 0.011 \\ 
1.213-1.408 & 1.309 & 1.161 & 0.015 & 0.012 \\ 
1.408-1.634 & 1.520 & 1.177 & 0.015 & 0.016 \\ 
1.634-1.896 & 1.762 & 1.186 & 0.014 & 0.016 \\ 
1.896-2.201 & 2.045 & 1.216 & 0.015 & 0.016 \\ 
2.201-2.555 & 2.373 & 1.208 & 0.014 & 0.015 \\ 
2.555-2.965 & 2.754 & 1.251 & 0.015 & 0.016 \\ 
2.965-3.441 & 3.195 & 1.243 & 0.015 & 0.015 \\ 
3.441-3.993 & 3.708 & 1.249 & 0.016 & 0.015 \\ 
3.993-4.635 & 4.302 & 1.296 & 0.017 & 0.015 \\ 
4.635-5.379 & 4.991 & 1.269 & 0.017 & 0.014 \\ 
5.379-6.243 & 5.793 & 1.296 & 0.019 & 0.014 \\ 
6.243-7.246 & 6.725 & 1.316 & 0.020 & 0.014 \\ 
7.246-8.409 & 7.801 & 1.303 & 0.022 & 0.014 \\ 
8.409-9.760 & 9.048 & 1.315 & 0.024 & 0.014 \\ 
9.760-11.327 & 10.499 & 1.272 & 0.025 & 0.013 \\ 
11.327-13.146 & 12.177 & 1.312 & 0.028 & 0.013 \\ 
13.146-15.258 & 14.146 & 1.225 & 0.029 & 0.012 \\ 
15.258-17.708 & 16.408 & 1.292 & 0.034 & 0.013 \\ 
17.708-20.552 & 19.044 & 1.299 & 0.037 & 0.013 \\ 

\hline
    \end{tabular}
  \end{center}
\end{table}

\begin{table}[hb]
  \caption
  [Muon Charge Ratio (Lynn Lake '97,'98,'99).]
  {Muon Charge Ratio (Lynn Lake '97,'98,'99).}
  \label{tab:muonratio99}
  \begin{center}
    \begin{tabular}{llccc}
      \hline
      \hline
  \multicolumn{5}{c}{Lynn Lake, Canada} \\
      \hline
Momentum & Mean  &    &      &       \\
Range & Momentum & $\mu^+ / \mu^-$ Ratio & Statistical Error & Systematic Error \\
(GeV/c) &  (GeV/c) &  &   &       \\
      \hline                                                            

0.576-0.901 & 0.739 & 1.217 & 0.023 & 0.000 \\ 
0.901-1.408 & 1.147 & 1.223 & 0.019 & 0.000 \\ 
1.408-2.201 & 1.781 & 1.226 & 0.017 & 0.011 \\ 
2.201-3.441 & 2.761 & 1.274 & 0.018 & 0.009 \\ 
3.441-5.379 & 4.288 & 1.273 & 0.020 & 0.007 \\ 
5.379-8.409 & 6.676 & 1.328 & 0.025 & 0.005 \\ 
8.409-13.146 & 10.382 & 1.325 & 0.032 & 0.003 \\ 
13.146-20.552 & 16.170 & 1.388 & 0.044 & 0.001 \\ 

\hline
    \end{tabular}
  \end{center}
\end{table}

\renewcommand{\baselinestretch}{2.0}


\begin{thebibliography}{00}





\bibitem{FU98}
 Y.~Fukuda et al.,
 Phys. Rev. Lett. 81 (1998) 1562.
\bibitem{GHKL96}
T.~K.~Gaisser et al.,
Phys. Rev. D 54 (1996) 5578.
\bibitem{GH2002}
T.~K.~Gaisser and M.~Honda,
Preprint, hep-ph/0203272.
\bibitem{HON95}
 M.~Honda et al.,
 Phys. Rev. D 52 (1995) 4985 .
\bibitem{SA99}
 T.~Sanuki et al.,
 Ap. J. 545 (2000) 1135.
\bibitem{AL2000}
J.~Alcaraz et al.,
Phys. Lett. B 472 (2000) 215.
\bibitem{AJ99}
 Y.~Ajima et al.,
 Nucl. Instr. and Meth. A, 443 (2000) 71.
\bibitem{OR87}
 S.~Orito,
 KEK Report 87-19, Proceedings of the ASTROMAG Workshop,
 edited by J.~Nishimura, K.~Nakamura, A.~Yamamoto. (1987) p.111.
\bibitem{AY94}
 A.~Yamamoto et al., Adv. Space Res., 14 (1994) 75.
\bibitem{YA88}
 A.~Yamamoto et al.,
 IEEE Trans. Magn. 24 (1988) 1421.
\bibitem{agel} 
Y.~Asaoka et al., Nucl. Instr. and Meth. A, 416 (1998) 236.
\bibitem{tof97} 
Y.~Shikaze et al., Nucl. Instr. and Meth. A, 455 (2000) 596.
\bibitem{SheaSmart} M.~A.~Shea et al.,  Proc. 27th ICRC 2001(Hamburg), (2001) 4063.
\bibitem{GEOMAG} World Data Center for Geomagnetism, Kyoto University, 
\begin{verbatim}
	http://swdcdb.kugi.kyoto-u.ac.jp/trans/index.html
\end{verbatim}
\bibitem{brooke64} G.~Brooke and A.~W.~Wolfendale,
  Proc.~Phys.~Soc. 83 (1964) 843.
\bibitem{diggory74} I.~S.~Diggory et al.,
  J.~Phys.~A:Math.,Nucl.~Gen., 7 (1974) 741.
\bibitem{golden95} R.~L.~Golden et al., J.~Geophys.~Res., 100 (1995) 23,515.
\bibitem{sanuki-cerenkov} T.~Sanuki, KEK Proceedings 94-11 (1995) 53.
\bibitem{sagisaka86} S.~Sagisaka,
  IL NUOVO CIMENT 9C (1986) 809.
\bibitem{ADJ} Aerological Data of Japan, edited by Japan
  Meteorological Agency (monthly issued).
\bibitem{nasasp} NASA/Marshall Space Flight Center,
\begin{verbatim}
	http://wwwssl.msfc.nasa.gov/ssl/pad/solar/sunspots.htm
\end{verbatim}
\bibitem{climax} University of Chicago, Neutron Monitor Datasets,
\begin{verbatim}
	http://ulysses.uchicago.edu/NeutronMonitor/neutron_mon.html
\end{verbatim}
\bibitem{sanukid} Y.~Asaoka et al., Phys.~Rev.~Lett. 88 (2002) 051101-1.
\bibitem{HW62} P.~J.~Hayman {\it et al.}, Proc. Phys. Soc. London. 80 (1962) 710.
\bibitem{GD79} P.~J.~Green et al., Phys. Rev. D 20 (1979) 1598.
\bibitem{AB75} C.~A.~Ayre et al., J. Phys. G 1 (1975) 584.
\bibitem{NS72} B.~C.~Nandi et al., J. Phys. A 5 (1972) 1384.
\bibitem{AC71} O.~C.~Allkofer et al., Phys. Lett. B 36 (1971) 425.
\bibitem{BC71} B.~J.~Bateman et al., Phys. Lett. B 36 (1971) 144.
\bibitem{R84} B.~C.~Rastin, J. Phys. G 10 (1984) 1609.
\bibitem{TKO98} S.~Tsuji et al., J. Phys. G: Nucl. Part. Phys. 24 (1998)
	1805.
\bibitem{DMP93} M.~P.~DePascale et al., J. Geophysical Res. 98 (1993) 3501.
\bibitem{KBA99} J.~Kremer et al., Phys.~Rev.~Lett. 83 (1999) 4241.
\bibitem{AHR71} I.~C.~Appleton et al., Nucl. Phys. B 26 (1971) 365.
\bibitem{BHT75} J.~M.~Baxendale et al., J. Phys. G 7 (1975) 781.
\bibitem{NS722} B.~C.~Nandi et al., Nucl. Phys. B 40 (1972) 289.
\bibitem{R842} B.~C.~Rastin, J. Phys. G 10 (1984) 1629.

\end{thebibliography}
\end{document}